# Effect of gamma-strength on nuclear reaction calculations

*Igor* Kadenko[1], *Vladimir* Plujko[1,*], *Borys* Bondar[1], *Oleksandr* Gorbachenko[1], *Borys* Leshchenko[2], *Kateryna* Solodovnyk[1], *Oleksandr* Tkach[1], and *Viktor* Zheltonozhskyi[3].

[1]Nuclear Physics Department, Taras Shevchenko National University, Kyiv, Ukraine
[2]National Technical University of Ukraine "Kyiv Polytechnic Institute", Kyiv, Ukraine
[3]Nuclear Structure Department, Institute for Nuclear Research of NAS, Kyiv, Ukraine

**Abstract.** The results of the study of gamma-transition description in fast neutron capture and photofission are presented. Recent experimental data were used, namely, the spectrum of prompt gamma-rays in the energy range 2÷18 MeV from 14-MeV neutron capture in natural Ni and isomeric ratios in primary fragments of photofission of the isotopes of U, Np and Pu by bremsstrahlung with end-point energies $E_e$= 10.5, 12 and 18 MeV. The data are compared with the theoretical calculations performed within EMPIRE 3.2 and TALYS 1.6 codes. The mean value of angular momenta and their distributions were determined in the primary fragments $^{84}$Br, $^{97}$Nb, $^{90}$Rb, $^{131,133}$Te, $^{132}$Sb, $^{132,134}$I, $^{135}$Xe of photofissions. An impact of the characteristics of nuclear excited states on the calculation results is studied using different models for photon strength function and nuclear level density.

## 1 Introduction

Nuclear reactions with different projectiles provide an information on properties of the excited nuclear states and nuclear reaction mechanisms. They are also required to different applications. Specifically, data of (n,x γ) reactions with any outgoing particle (x) and gamma-rays for atomic nuclei of the reactor materials and photofission reactions are needed for estimations of energy release, γ-ray shielding and radiation swelling of the reactor pressure vessel internals. Photofission reactions are also essential to an explanation of nuclear fission dynamics.

Here, we consider reliability of description of gamma-transitions in (n,x γ) reactions and photofission by the use of simple closed-form expressions [1] for photon strength function (PSF) as well as for nuclear level density (NLD). We use our recent experimental data, namely, the spectrum of prompt gamma-rays in the energy range 2÷20 MeV from 14 MeV neutron capture of $^{nat}$Ni and isomeric ratios in primary fragments of photofission of isotopes U, Np and Pu by bremsstrahlung with end-point energies $E_e$ = 10.5, 12.0 and 18.0 MeV. These data are compared with the theoretical calculations performed within EMPIRE 3.2 and TALYS 1.4 codes [2,3]. An impact of shapes of electric dipole PSF and NLD on accuracy of cross section description and determination of mean angular momenta in primary fragments is analyzed.

## 2 Experimental data

Figure 1 demonstrates comparison of our recent experimental data [4] for γ-spectrum from (n,x γ) reactions on $^{nat}$Ni with the earlier data from EXFOR database [5-8].

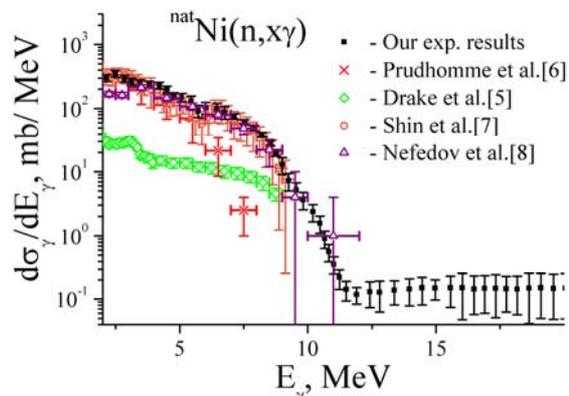

**Fig. 1.** The measured gamma spectrum from fast neutron capture in $^{nat}$Ni in comparison with experimental data from Refs.[5-8].

In our experiment, the amplitude spectra of gamma-rays were measured in a circular geometry using time-of-flight technique based on pulse neutron generator of neutrons with energy 14.1 MeV (NPG-200), designed and manufactured at Nuclear Physics Department of Taras Shevchenko National University of Kyiv (Ukraine). All experiment and calculation details can be found in ref. [4,9].

One can see in Fig.1, that for the gamma-ray energy interval 2÷6 MeV, our experimental results are in rather close agreement with the data from EXFOR database. Gamma-ray spectrum smoothly decreases with gamma-ray energy. For higher energy range of 12÷18 MeV, the differential cross section $d\sigma_\gamma(E_\gamma)/dE_\gamma$ has almost constant value.

In this study of PSF and NLD, we use also our recent experimental data [10,11] for isomeric yield ratios in primary fragments of photofission of isotopes U, Np and Pu by bremsstrahlung with different end-point energies $E_e$.

**Table 1.** Isomeric yield ratios and mean angular momentum $\overline{J}$ of photofission primary fragments calculated within EMPIRE 3.2 code with default inputs

| Fragment | Target nucleus | Energy $E_e$,MeV | Isomeric ratio ($Y_m/Y_g$) | $\overline{J}/\hbar$ |
|---|---|---|---|---|
| $^{84}$Br | $^{235}$U | 18 | 0.14±0.01 | 1.8±0.5 |
| $^{84}$Br | $^{237}$Np | 18 | 0.15±0.01 | 1.9±0.5 |
| $^{84}$Br | $^{239}$Pu | 18 | 0.118±0.006 | 1.7±0.5 |
| $^{90}$Rb | $^{237}$Np | 18 | 1.2±0.2 | 2.2±0.6 |
| $^{90}$Rb | $^{239}$Pu | 18 | 1.0±0.2 | 1.8±0.6 |
| $^{97}$Nb | $^{235}$U | 10.5 | 0.7±0.09 | 1.4±0.6 |
| $^{97}$Nb | $^{238}$U | 12 | 0.75±0.09 | 1.4±0.6 |
| $^{97}$Nb | $^{238}$U | 18 | 3.8±0.6 | 5.0±0.7 |
| $^{131}$Te | $^{235}$U | 18 | 2.6±0.5 | 6.8±0.8 |
| $^{131}$Te | $^{237}$Np | 18 | 1.9±0.3 | 5.8±0.7 |
| $^{131}$Te | $^{239}$Pu | 18 | 3.2±0.6 | 7.4±0.8 |
| $^{132}$Sb | $^{235}$U | 18 | 1.46±0.22 | 8.0±0.7 |
| $^{132}$Sb | $^{237}$Np | 18 | 1.01±0.12 | 6.9±0.6 |
| $^{132}$Sb | $^{239}$Pu | 18 | 1.48±0.16 | 8.1±0.6 |
| $^{132}$I | $^{235}$U | 18 | 2.2 ±0.4 | 9.6±0.9 |
| $^{132}$I | $^{237}$Np | 18 | 0.95± 0.15 | 6.7±0.7 |
| $^{132}$I | $^{239}$Pu | 18 | 0.51± 0.06 | 5.2±0.6 |
| $^{133}$Te | $^{235}$U | 18 | 4.3± 0.3 | 7.6±0.6 |
| $^{133}$Te | $^{237}$Np | 18 | 9.0± 0.9 | 10.6±0.7 |
| $^{133}$Te | $^{239}$Pu | 18 | 5.3± 0.3 | 8.4±0.5 |
| $^{134}$I | $^{235}$U | 18 | 0.58± 0.09 | 5.6±0.6 |
| $^{134}$I | $^{239}$Pu | 18 | 1.26±0.25 | 7.7±0.8 |
| $^{135}$Xe | $^{235}$U | 18 | 0.056±0.007 | 1.4±0.5 |
| $^{135}$Xe | $^{237}$Np | 18 | 0.041±0.006 | 1.2±0.5 |
| $^{135}$Xe | $^{239}$Pu | 18 | 0.066±0.007 | 1.4±0.5 |

These measurements were performed using the bremsstrahlung induced by electron beam from the M-30 microtron at the Laboratory of Photonuclear Reactions of the Institute of Electron Physics (Uzhgorod, Ukraine). The activation method was applied with direct spectrometry of irradiated samples for further identification of the radioactive products. Isomeric ratios were calculated as ratio of the yields of the reactions, leading to formation of final nuclei in the metastable and ground states $R_Y = Y_m/Y_g$. All other details can be found in [10-12]. The results are presented in Table 1.

# 3 Theoretical calculations

Theoretical calculations of cross-sections and mean angular momentum are performed using EMPIRE 3.2 and TALYS 1.6 codes. Comparison with experimental data is made in two steps. At first we use estimations obtained with default parameters and then we use results of calculations within EMPIRE code with different shapes of PSF and NLD [13-15].

Default expressions for PSF and NLD are the following ones: MLO1 variant of the Modified Lorentzian model (MLO) for the electric dipole PSF with Enhanced Generalized Super-Fluid Model (EGSM) for NLD in Empire code, and Enhanced Generalized Lorentzian (EGLO) for PSF with Gilbert-Cameron approach (GC) for NLD in TALYS code. Calculations were performed with allowance for outgoing particles and gamma-rays at equilibrium (HF denotations in the figures) and from pre-equilibrium states (HF+PE denotations with parameter PCROSS = 1.5 for EMPIRE). Global optical potential given by Koning – Delaroche [16] was adopted as default in the calculations within two codes.

For calculations of considered characteristics, we use expressions presented in Refs.[4, 11-12]. The cross section of the target with natural elements was taken as a sum of the cross-sections for each isotope of the target weighted with the abundances of the isotope, and the cross section of gamma-ray emission for isotope was a sum of the cross-sections for all possible reactions with any outgoing particle and gamma-rays.

Measured isomeric ratio $R_Y$ for primary fission fragment $(A_f, Z_f)$ was used to obtain its mean angular momentum $\overline{J}$. Initially the spin distribution of primary fission fragments may be deduced from theoretical and experimental isomeric ratios comparison. The theoretical values of the isomeric ratios are determined by the generalized Huizenga-Vandenbosh statistical model [10-12]. Populations of ground and metastable states of given nuclide $(A_f, Z_f)$ after decaying of $(A_f + i, Z_f)$ isotope with larger number of neutrons ($i \le i_m = 2$) were also taken into account. The probabilities of populations of ground and metastable states by cascades of gamma-rays and neutrons were calculated using EMPIRE 3.2 and TALYS 1.6 codes. The probabilities of deexcitation and population of discrete levels were taken from the RIPL-3 library [1].

Figure 2 shows comparison between experimental data and calculations within EMPIRE and TALYS codes of cross-section for (n,x $\gamma$ ) reaction on $^{nat}$Ni.

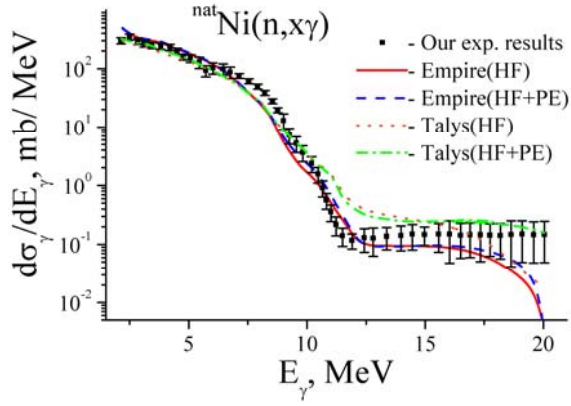

**Fig. 2.** Comparison of the experimental and theoretical gamma spectrum from fast neutron capture in $^{nat}$Ni. Calculations were performed using EMPIRE and TALYS codes with default parameters

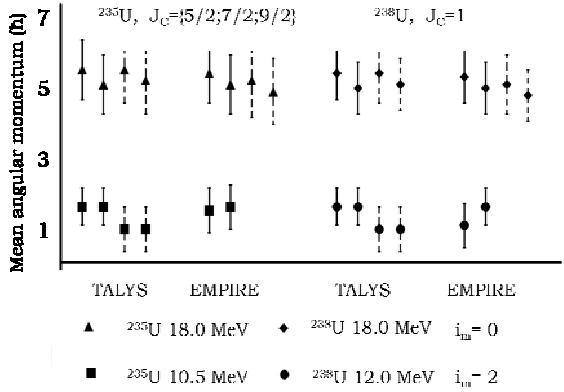

**Fig. 3.** Comparison of theoretical calculations of mean angular momentum of $^{97}$Nb performed using EMPIRE and TALYS codes with default parameters

It can be seen from Fig. 2 that in the gamma-ray energy interval 2÷10 MeV the results of calculations are in rather good agreement with experimental data and a role of preequilibrium emission is small. For high energy range 10÷20 MeV, contribution of the preequilibrium processes is important.

Table 1 shows the values (in a unit of $\hbar$) of mean angular momentum $\overline{J}$ obtained with default parameters for primary fragments $^{84}$Br, $^{90}$Rb, $^{97}$Nb, $^{131,133}$Te, $^{132}$Sb, $^{132,134}$I, $^{135}$Xe of photofission by bremsstrahlung. Populations were calculated within EMPIRE code. Figure 3 demonstrates comparison between values of mean angular momentum obtained within EMPIRE and TALYS codes for $^{97}$Nb in photofission of $^{235}$U and $^{238}$U by bremsstrahlung with end-point energy of 10.5, 12 and 18 MeV. One can see that calculation results with use of different codes are in close agreement.

Figures 4-7 demonstrate results of calculations within EMPIRE code (PCROSS=1.5) with different shapes of PSF and NLD[13-15]. For the PSF, we used the EGLO model, MLO1 and MLO4 variants of MLO approach, Standard Lorentzian model (SLO) and Generalized Fermi Liquid (GFL) model. For the

NLD, we applied the EGSM, Generalized Superfluid Model (GSM), Gilbert and Cameron (CG) model (from EMPIRE 2.18), microscopic combinatorial level densities within Hartree-Fock-Bogoliubov method (HFBM) and Modified Generalized Super-Fluid Model with Bose attenuated numbers for vibrational enhancement factor (MEGSM) [1,15].

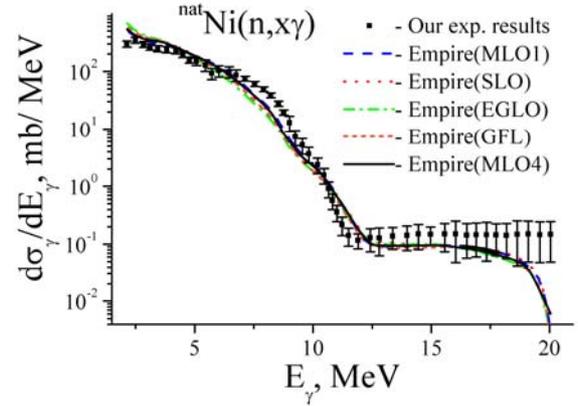

**Fig. 4.** The gamma spectrum from fast neutron capture in $^{nat}$Ni calculated within EMPIRE 3.2 code with different models for the PSF

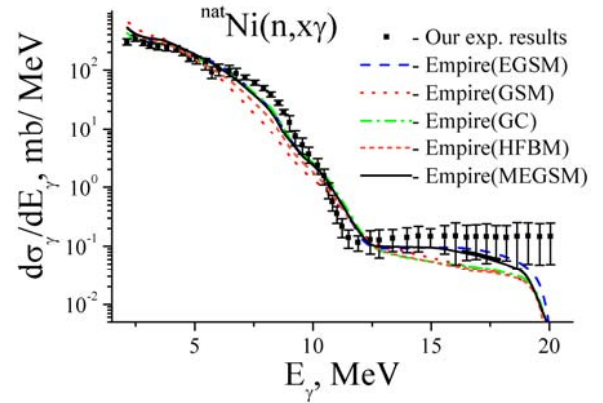

**Fig. 5.** The gamma spectrum from fast neutron capture in $^{nat}$Ni calculated within EMPIRE 3.2 code with different models for NLD

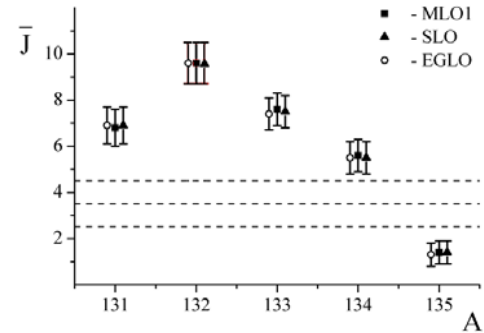

**Fig. 6.** Mean angular momentum of primary fragments $^{131}$Te, $^{132}$I, $^{133}$Te, $^{134}$I, $^{135}$Xe calculated by EMPIRE 3.2 code with different PSF models

One can see from Figs. 4,5 that difference between calculated gamma spectra with different PSF and NLD shapes is small. The best agreement with the experiment was obtained in case of using MLO1 or MLO4 for the PSF and EGSM or MEGSM for the NLD.

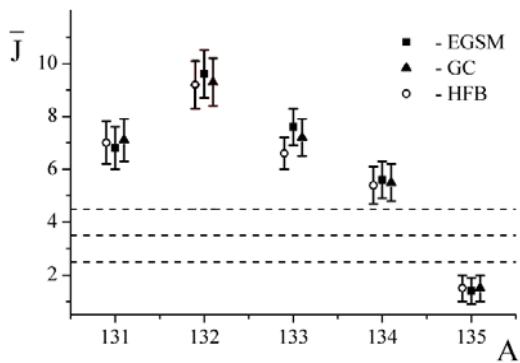

**Fig. 7.** Mean angular momentum of primary fragments of $^{131}$Te, $^{132}$I, $^{133}$Te, $^{134}$I, $^{135}$Xe calculated with EMPIRE 3.2 code with different shapes of NLD.

Figures 6,7 show that values of mean angular momentum of primary fission fragments are unaffected by PSF and NDL models.

## 4 Conclusions

The theoretical calculations of gamma-spectrum from (n,x γ) reaction and mean angular momentum of primary fragments were performed. New experimental data were used for (n,x γ) cross-section on $^{nat}$Ni induced by 14 MeV neutrons and for isomeric ratios in primary photofission fragments ($^{84}$Br, $^{90}$Rb, $^{97}$Nb, $^{131,133}$Te, $^{132}$Sb, $^{132,134}$I and $^{135}$Xe ).

For two different quantities, i.e. cross-sections and mean angular momentum, the comparison between calculations within EMPIRE and TALYS codes shows small impact of shapes of RSF and NLD on the results.

Comparisons of the experimental data with their theoretical values also demonstrate high reliability of the calculations with the use of the EMPIRE and TALYS codes with the default sets of the input parameters for estimations of both the gamma-ray spectrum in reactions induced by fast neutrons and determination of mean angular momentum in primary fission fragments.

This work is supported in part by the IAEA (Vienna) under IAEA Research Contract within CRP #F41032